\documentclass{amsart}
\usepackage{graphicx,amssymb}
\begin{document}
\def\eq#1{{eq.~(\ref{#1})}}
\def\eqs#1#2{{eqs.~(\ref{#1})--(\ref{#2})}}
\newcommand{\nn}{\nonumber}
\newcommand{\cal}{\mathcal}
\newcommand{\cy}{{\cal Y}(\theta)}
\newcommand{\cx}{{\cal X}(\theta)}
%
%
%
%
\title{Excited-state energies and supersymmetric indices}

\author{Paul Fendley}
\address{
Los Alamos National Laboratory\\
Theoretical Division T-8, MS B285\\
Los Alamos, NM 87544. Address after Aug.\ 25, 1997:\\
Physics Dept., University of Virginia\\
Charlottesville, VA 22901}

\begin{abstract}
We study multi-soliton states in two-dimensional
$N$=2 supersymmetric theories.
We calculate their energy exactly as a
function of mass and volume in the simplest integrable $N$=2
supersymmetric theory, the sine-Gordon model at
a particular coupling.  These energies are related to the
expectation value ${\cal I}={\rm tr} [\exp(in\pi F) \exp(-H/T)]$, where
$F$ is the fermion number. For $n$=1, this is Witten's index; for $n$
an odd integer, we argue that ${\cal I}$ is an index in the sense that
it is independent of all $D$-term variations.
\end{abstract}

\maketitle

\section{Introduction}	

The thermodynamic Bethe ansatz (TBA) has proven a useful tool for
computing the ground-state (Casimir) energy of an integrable $1+1$
dimensional system \cite{YY}.
The special properties implied by integrability allow one to compute the
partition function at any temperature $T$
\begin{equation}
\label{forz}
Z\equiv {\rm tr}\left[e^{-H_L/T}\right]
\end{equation}
where $H_L$ is the Hamiltonian for the system on a line
of length $L$. The TBA calculation gives the free energy per unit
length
$-T \ln Z/L$ when $L$ much is
larger than any other length scale in the
theory.
This partition function is related to the ground-state energy by interchanging
the definition of space and time.
The above trace is equivalent to a Euclidean path integral where space
time is a cylinder of length $L$ and circumference $1/T$. This path
integral can equivalently by computed by taking space to be a circle of
circumference $R\equiv 1/T$, and time propagation to be in the $L$-direction:
\begin{equation}
Z = {\rm tr}\left[e^{-H_R L}\right]
\label{modz}
\end{equation}
Since $L$ is very large, the leading contribution to this trace is from the
lowest eigenvalue of $H_R$, which is the ground-state energy
$E_0$. Thus
\begin{equation}
E_0=- {1\over L} \ln Z 
\label{forE}
\end{equation}
This trick of reversing space and time is known
in the conformal field theory literature as a modular
transformation.

The TBA can be generalized in order to calculate excited-state
energies as well. Adding imaginary chemical potentials to (\ref{forz})
corresponds to changing the boundary conditions around the cylinder
(i.e.\ putting some sort of twist in the $R$-direction). This enables
one to project out the ground state \cite{cardy} and
calculate excited-state energies \cite{me}. In this manner one can
calculate a few low-lying excited-state energies in a number of models
\cite{MKM,me}, but all those calculated here
have the property that for large $R$, they are degenerate with the ground
state and thus not multi-particle states.

Recently, several techniques were developed which allow the calculation
of the energies of multi-particle states \cite{BLZ,dorey}.
The basic idea is to
analytically continue the TBA equations onto another sheet in an
appropriate space (the fugacity in \cite{BLZ}, the mass in
\cite{dorey}). Using formal techniques developed in \cite{KP}, the
results of the continuation can be written in closed from, and are
interpreted as the TBA equations for an excited state. These
generalized TBA equations have been worked out in detail only for
several states in the Yang-Lee model \cite{BLZ,dorey} and some
generalizations \cite{doreynew}.

We give explicit thermodynamic Bethe ansatz equations for
excited-state
energies in a unitary theory, the simplest integrable $N$=2
supersymmetric field theory.
In this model, which is equivalent to the ordinary
sine-Gordon model at a special value of the coupling, we calculate
an infinite number of eigenvalues of the Hamiltonian, namely
\begin{equation}
E_n=\langle n| H_R |n\rangle
\label{eexp}
\end{equation}
when $n$ is a multi-soliton states on a circle of circumference $R$.
The energy $E_n$ of this excited state is a function of the volume $R$ of
the one-dimensional space and the mass
scale $m$ of the theory. The ground-state energy $E_0$ is
the Casimir energy on this circle, and it vanishes as
$R\to\infty$. The excited states in general are a collection of
interacting particles, so as $R\to\infty$, $E_n\to am$, with
$a$ depending on how many particles are in the state and their masses.
If there is only one particle mass in the theory, then $a$ is an integer.

At first glance, the supersymmetric model to be discussed here
is more complicated than the Yang-Lee model studied in detail in
\cite{BLZ,dorey}. It has two particles in the spectrum
(the soliton and antisoliton in sine-Gordon)
with a non-diagonal $S$-matrix
instead of a single particle with a single $S$-matrix element.
However, because of various simplifying features
probably related to the supersymmetry, we are able to find the
appropriate equations to describe an infinite number of excited-state
energies. The excitations in this theory are all solitons saturating
the Bogomolny bound \cite{pk}, so
our results give the exact non-perturbative energy of an infinite
hierarchy of interacting
multi-soliton states.

Some of these excited-state energies have a particular interest in
the context of $N$=2 theories.
Formally, one can view the calculation in terms of
\begin{equation}
{\cal I}(\alpha)\equiv {\rm tr} \left[e^{i\alpha F}e^{-R H_L}\right],
\label{fun}
\end{equation}
corresponding to adding imaginary chemical potentials proportional
to the fermion number $F$.
${\cal I}(\alpha)$ makes sense as a function of $\alpha$ at least for
$\alpha$ small, and is real when the theory is $CP$-invariant.
It depends on $R,L$ and any mass scale $m$ in the theory.
By analogy with (\ref{forE}) we
define
\begin{equation}
\ln {\cal I}(\alpha)=-E(\alpha)L
\label{EI}
\end{equation}
for large $L$. Below we discuss how to define $E(\alpha)$ for all $\alpha$.
The point of \cite{BLZ} is that at
appropriately-chosen values $\alpha_n$, one has
\begin{equation}
E_n=E(\alpha_n)
\end{equation}
In this supersymmetric case, we study $\alpha_n=n\pi$. 
We show that when appropriately defined, ${\cal I}(\alpha)$ is {\bf not}
periodic in $\alpha$, even though $F$ is half-integer in
our example. (This arises
because of subtleties in taking the infinite-space limit, as are common
in open-space index theorems \cite{CBS}.) 

In supersymmetric theories, an object of
fundamental importance is Witten's index \cite{Witten}
\begin{equation}
{\cal I}(\pi)={\rm tr}\left[(-1)^F e^{-H_L/T}\right]
\label{witten}
\end{equation}
In the case discussed here,  Witten's index is independent of $R,L$ and $m$. 
Another case of interest in $N$=2 theories is
$dI(\alpha)/d\alpha|_{\alpha=\pi}$, which is the ``new supersymmetric index''
calculated in \cite{CFIV}. This is an index in the sense that it is
independent of all $D$-term perturbations of the theory. In this paper,
we will show that ${\cal I}(n\pi)$ is also independent of the
$D$-term when $n$ is an an odd integer.

More precisely, in section 2 we show that ${\cal I}(n\pi)$ for $n$
an odd integer corresponds to computing Witten's index with different
open-space asymptotic conditions. These asymptotic conditions
correspond to inserting operators at the ends of the space-time
cylinder. We argue that these objects are independent of
$D$-term type perturbations, and thus are open-space
indices in the sense of \cite{CFIV}. We also show the ${\cal
I}(\alpha)$ gives excited-state energies when $n$ is any
integer. In section 3 we use the thermodynamic Bethe ansatz to derive
equations for ${\cal I}(\alpha)$ in our example. In section 4, we
discuss the solutions in the $m\to 0$ limit, where ${\cal I}(\alpha)$ is
an analytic function of $\alpha$. In section 5 we extend the
results to all $mR$, showing how to define ${\cal I}(\alpha)$ past
discontinuities at certain values of $\alpha$. In the first appendix, we
check our results by explicitly computing the energy in the massless limit.
In the second appendix, we calculate the excited-state energies of
a Dirac fermion, mainly in order to clarify the nature of the discontinuity
arising in section 5.

\section{Supersymmetric indices and excited states}

In this section we explain how the excited-state energies discussed
in the introduction are related to interesting quantities in $N$=2
supersymmetric field theories.

In these two-dimensional models,  Euclidean spacetime is a cylinder of
circumference $R$ and very long length $L$.  Varying $\alpha$ in
(\ref{fun}) corresponds to changing boundary conditions as one goes
around the cylinder. In particular, for $\alpha=n\pi$, when $n$ is an
even integer these correspond to antiperiodic boundary conditions on
fermionic operators. When $n$ is an odd integer, these are periodic
boundary conditions on the fermions. In conformal field theory (which
applies when $m\to 0$), these are called Neveu-Schwarz and Ramond
boundary conditions respectively.

The energy $E(\alpha)$ in (\ref{EI}) is a natural object to consider
when one thinks of space as in the $R$-direction, and time in the
$L$-direction. It is a particular eigenvalue of the Hamiltonian $H_R$
operating on the space of states  $|\,\rangle_{\alpha}$ on a circle of
circumference $R$. The Hilbert space is labeled by $\alpha$ because
changing the boundary conditions around $R$ changes the states allowed
in the Hilbert space.  Shifting $\alpha$ by $2\pi$ takes
the space of states $|\,\rangle_{\alpha}$ to itself,
so for example the Ramond sector is
mapped to the Ramond sector. However, it does not take a given state to
itself. This is well known in the superconformal literature where the
change in $\alpha$ is often called {\bf spectral flow}; see e.g.\
\cite{SS}. Thus in the $m\to 0$ limit, $E_n$ for $n$ odd is the energy
of some state in the Ramond sector, while for $n$ even it is the energy
of some state in the Neveu-Schwarz sector.

In a conformally-invariant
theory, the energy of a state must be proportional to $1/R$
because there is no other scale in the theory. The constant of
proportionality is related to the central charge $c$ of the theory and
the left and right conformal dimensions ($h_n,\overline h_n$) of the operator
which creates the state $|n\rangle$ \cite{BCNA}:
\begin{equation}
\lim_{m\to 0} E_n = -{\pi\over 6R}(c-12h_n-12\overline h_n).
\label{Ec}
\end{equation}
Thus in the $m\to 0$ limit, the ground-state energy in the conformal limit
is related to the central charge, while excited-state energies are
related to the dimensions of the operators which create these states.
Thus when $n$ is an even (odd) integer, we can identify which operator in
the Neveu-Schwarz (Ramond) sector created the state.

In an $N$=2 theory, spectral flow arguments give \cite{SS}
\begin{equation}
\lim_{m\to 0} E(\alpha)
= -{\pi c\over 6R}\left(1-\left({\alpha\over\pi}\right)^2\right).
\label{Eal}
\end{equation}
Since
calculations here are left-right symmetric,
the states $|n\rangle$ are therefore created by operators with
conformal dimension $(h_n,\overline h_n)=(cn^2/24,cn^2/24)$.
The formula (\ref{Eal}) is well known
in the example we will discuss in later sections,
the sine-Gordon model at its $N$=2 supersymmetric point.
The $m\to 0$ limit consists of a free boson, which has $c=1$.
\footnote{Conformal
field theory aficionados will notice that (\ref{Eal}) looks like the
central charge of the minimal models. A non-zero $\alpha$ corresponds
to a non-zero charge at infinity in the Coulomb-gas picture
\cite{NienDF}. Taking $\alpha=\pi$ gives the trivial $c=0$ minimal
model.}

Notice that at $n=1$, we have $E_1=0$. In fact, $E_1=0$ for any $m$, not just in the conformal limit (see \cite{CFIV} for
the TBA calculation in many $N$=2 theories). This is a deep result
for supersymmetric
theories, because from the point of view where space is in the
$L$-direction and time is in the $R$-direction, we are calculating the
leading term in Witten's index \cite{Witten}. Witten's result is that
if there are periodic boundary conditions in the $L$-direction, then
${\cal I}(\pi)$ in (\ref{witten}) is related to the number of
ground states,
and is an integer independent of $L, T$ and $m$. Thus when we find
$E$ by interchanging the roles of space and time, (\ref{EI}) shows that
$E_1=0$ because ${\cal I}$ is finite and $L\to\infty$. 
With periodic boundary conditions in the $L$-direction,
Witten's index is independent of any supersymmetry-preserving
deformations of the theory.

In general ${\cal I}(\alpha)$ does depend on
perturbations, for example, changing the mass.
However, when $n$ is odd, ${\cal I}(n\pi)$ does have a special property.
We argue here that it is independent of any $D$-term perturbations.
There are four supercharges $Q^+,Q^-,\overline Q^+,\overline Q^-$ in an $N$=2
theory (for any $m$). Both $Q^+$ and $\overline Q^-$ have
fermion number $+1$, while the other two have fermion
number $-1$. These charges are defined with periodic boundary
conditions, so for the rest of this section we restrict to the case $n$
an odd integer. In a superspace action a $D$-term comes from
integrations over all four Grassman coordinates. This means that a
variation of a $D$-term can be written as inserting $\{ Q^+, [\overline
Q^-,\Lambda (x) ]\} $ in the path integral where $\Lambda$ itself can
be written as $\{ Q^-, [\overline Q^+ , K]\}$ for some $K$. Thus a $D$-term
variation of ${\cal I}(n\pi)$ can be written as
\begin{equation}
\delta {\cal I}(n\pi)= {1\over L}\int dx\,\langle n|
\{ Q^+, [\overline Q^-,\Lambda (x) ]\} e^{-H_R L}|n\rangle
\label{vary}
\end{equation}
Because $Q^+$ and $\overline Q^-$ anticommute
with each other, if either $Q^+|n\rangle=0$ or $\overline Q^-|n\rangle=0$,
then $\delta I=0$.

In a superconformal field theory, a state annihilated by $Q_+$ is taken
to another one annihilated by $Q_+$ under spectral flow by positive
$\alpha$. We prove this by writing the left-moving and
right-moving fermion currents $J(x+t)$ and $\overline J(x-t)$
in terms of a free boson: $J=\delta_x\Phi(x,t)$, and
$\overline J=-\delta_x\overline \Phi(x,t)$.
The fermion number $F=\int_0^R dx [J(x+t)-J(x-t)]$. In the sine-Gordon
example to be
discussed below, $\Phi-\overline\Phi$ is just the $c=1$ boson,
normalized so that the perturbing operator of dimension ($2/3,2/3$)
is $\cos[2(\Phi-\overline\Phi)]$.
Then $Q^+=\int dx A\exp(i3\Phi(x,t)/c)$, while $\overline
Q^-=\int dx B
\exp(i3\overline\Phi(x,t)/c)$, where $A$ and $B$ commute with
$\Phi$ and $\overline\Phi$. The exponents can be verified
by checking that they give the supercharges the correct fermion number.
The operator which implements the spectral flow can be written in terms
of this boson.
The state $|n\rangle$ in the Ramond sector is \cite{LVW}
$$|n\rangle = \int_0^R dx\, e^{is(\Phi(x,t)-\overline\Phi(x,t))}
|1\rangle.$$
where $s=(n-1)/2$ is the number of ``units'' of spectral flow; one
unit of spectral flow corresponds to shifting $\alpha$ by $2\pi$.
Therefore, if $s$ is positive, the spectral flow
operator commutes with $Q^+$, because the exponents of both are
positive. If $s$ is negative, the spectral flow commutes with
$\overline Q^-$. Since the state $|n=1\rangle$ yields Witten's
index, it is annihilated by $Q^+$ and $\overline Q^-$ (and the other
supercharges as well). Therefore, the states with $n$ an odd integer
are all annihilated by either $Q^+$ or $\overline Q^-$.
This argument applies only at the superconformal point. To go
off the critical point while preserving supersymmetry, one adds to the
Hamiltonian a term which commutes with all the supercharges. Therefore,
in particular, it commutes with $Q^+$, and $\delta I$ in
(\ref{vary}) is still zero.

We have therefore shown that the quantities ${\cal I}(n\pi)$ are
independent of any $D$-term variation of the theories when $n$ is an
odd integer. Generically, there are only a finite number of
perturbations of a theory which are not $D$-term, so this means ${\cal
I}(n\pi)$ depends on only a finite number of parameters. In the example we
will discuss below, there is only one.

One can think of ${\cal I}(n\pi)$ with $n$ odd as an open-space index.
Before Witten's work, in fact, it was shown that there are theories
where (\ref{witten}) is not an integer and does depend on various
parameters \cite{CBS}.  The arguments of \cite{Witten} utilize
periodic conditions and hence do not necessarily apply in open space.
The expectation value (\ref{witten}) can vary when defined on
an open space with appropriate asymptotic conditions. We have shown
that ${\cal I}(n\pi)$ with $n$ odd amounts to calculating
(\ref{witten}) with asymptotic conditions corresponding to
inserting the state $|n\rangle$ at the end of the cylinder. For the
subsequent analysis, we do not need to know how to write these
boundary conditions in terms of the fields in $H_L$. Knowing how would
be interesting, in order to compare our results with \cite{CBS} and
subsequent open-space index results like \cite{poly}.

${\cal I}(\alpha)$ as defined in (\ref{fun}) seems
periodic in $\alpha$ when $F$ is rational.
Periodicity in $\alpha$ contradicts for example the result (\ref{Eal}).
Thus the definition (\ref{fun}) should be understood as the appropriate
continuation away from $\alpha=0$. In particular, below we will
calculate $E(\alpha)$ for all $\alpha$ and use this to define ${\cal
I}(\alpha)$. As with the open-space index theorems, this seems to be
the only way to precisely define ${\cal I}(\alpha)$ for all $\alpha$
when $L\to\infty$.

\section{The thermodynamic Bethe ansatz equations}

In the previous section we showed that $E(\alpha)=-\ln {\cal
I}(\alpha)/L$ gives excited-state energies in $N$=2 supersymmetric
theories when $\alpha=n\pi$. For the remainder of this paper, we show
how to calculate $E_n\equiv E(n\pi)$ for the sine-Gordon model at its $N=2$
supersymmetric point. This is the first model in the $N$=2 discrete
series perturbed by its only relevant supersymmetry-preserving operator
(see e.g.\ \cite{pk} for a discussion of these models). This model is
interesting not just because of its supersymmetry. Because it
corresponds to sine-Gordon coupling  $\beta^2=16\pi/3$ (in the usual
normalization where the kinetic term in the action has a $1/2$ in
front, and the free fermion point coupling $\beta^2 =4\pi$), it is in
the ``repulsive'' regime, where the fermions in the related Thirring
model repel each other. The cases discussed in
\cite{BLZ,dorey,doreynew} are all in the attractive regime. Moreover,
as we mention in the conclusion, our model gives information on 2d
circular polymers.

Our starting point is the thermodynamic Bethe ansatz (TBA) equations in
functional form. Here, these equations are
written in terms of two entire functions $X(\theta;\alpha)$ and
$Y(\theta;\alpha)$
\begin{eqnarray}
X(\theta+i\pi/2) X(\theta-i\pi/2) &=&1+2\cos\alpha \, Y(\theta) +Y^2(\theta)\\
Y(\theta+i\pi/2) Y(\theta-i\pi/2)&=& 1+ X(\theta)
\label{tba}
\end{eqnarray}
where for ease of notation we often don't write out the $\alpha$
argument. For $\alpha\le \pi$ these equations can be derived from the
underlying lattice model \cite{TS,Fowler} or directly from the
quasiparticle $S$ matrix \cite{pk,CFIV}. It follows from
(\ref{tba}) that when $X\ne 0$, $Y(\theta)=Y(\theta+3i\pi)$ and
$X(\theta)=X(\theta+3i\pi)$. Even though the equations are periodic in
$\alpha$, there is no guarantee that the solutions will be.

To relate these functions to the energy levels of the system, we
utilize a conjecture of \cite{BLZ}. This conjecture relates the TBA
quantities to the infinite number of conserved quantities $I_n(\alpha)$ in an
integrable theory.
In this case ($\beta^2=2/3$ in their conventions),
their conjecture relates the asymptotic expansion of $Y(\theta;\alpha)$
around $\theta=-\infty$ to the $I_n$:
\begin{equation}
\ln Y(\theta;\alpha) = \sum_{n=1}^{\infty}C_n e^{(2n-1)\theta}
I_{2n-1}(\alpha)
\label{asyexp}
\end{equation}
where the the convention-dependent constants $C_n$ are proportional to
$R^{2n-1}$. The first conserved quantity $I_1(\alpha)$ is just the energy
$E(\alpha)$. As we will see momentarily, this is equal to the ground-state
energy determined by standard TBA methods.

 To make further contact with the physics, we note that $\theta$ is the
rapidity of the quasiparticles of the problem, defined so that when the
quasiparticles have mass $m$, the energy of the particle is
$m\cosh\theta$ and the momentum is $m\sinh\theta$. The filling
fraction (the Fermi distribution function generalized to this
interacting problem) at temperature $T\equiv 1/R$ is $X/(1+X)$ when
$\alpha\le\pi$ \cite{pk}. This means that the probability that there is a
particle with rapidity between $\theta$ and $\theta+d\theta$ is
$d\theta X(\theta)/(1+X(\theta))$. At high energy (large $\theta$) the
particle interactions are irrelevant, so we can treat these particles
as free and having a Fermi distribution function. This implies an
asymptotic condition for $X(\theta)$:
\begin{equation}
X(\theta) \to e^{-mR\cosh\theta}\qquad\hbox{for }\theta\to\pm\infty
\label{asy}
\end{equation}
which in turn requires
\begin{equation}
Y(\theta) \to 1 \qquad\hbox{for }Re\theta\to\pm\infty,\ |Im\theta|<\pi/2
\label{asyii}
\end{equation}

Even with the asymptotic condition (\ref{asy}),
there are an infinite number of solutions to (\ref{tba}). To fix a unique
solution, we utilize a lemma of \cite{KP}:

\bigskip
\noindent
{\bf Lemma}: When $f(\theta)$ is analytic
and bounded in the strip
$-\pi/2 \leq Im(\theta) \leq \pi/2$ and obeys the relation
$$f(\theta+i\pi/2) +f(\theta-i\pi/2)= g(\theta)$$
for some $g(\theta)$, then
\begin{equation}
f(\theta)=\int_{-\infty}^{\infty} {d\theta'\over 2\pi} {1\over\cosh(\theta-\theta')}
g(\theta')
\end{equation}

\noindent
Another derivation of this lemma is given in \cite{TW}.

If there are
no zeroes in $X(\theta)$ or $Y(\theta)$ in $|Im\theta|
\le \pi/2$ and $X$ obeys the asymptotic condition (\ref{asy}), then
both $\ln[X(\theta)e^{mR\cosh\theta}]$ and $\ln Y(\theta)$ satisfy
the conditions of the lemma above. Thus there is a unique solution
$X_0(\theta), Y_0(\theta)$
of (\ref{tba}) with no zeros and which satisfies the asymptotic condition.
The lemma requires that they solve the integral equations:
\begin{eqnarray}
\quad\ln X_0(\theta)&=&-mR\cosh\theta
\nonumber\\
&&\ + \int {d\theta'\over 2\pi}
{1\over\cosh(\theta-\theta')}\ln\left[(1+e^{i\alpha}Y_0(\theta'))
(1+e^{-i\alpha}Y_0(\theta'))\right]\quad
\label{tbanozeroi}\\
\quad\ln Y_0(\theta)&=& \int {d\theta'\over 2\pi}
{1\over\cosh(\theta-\theta')}\ln(1+X_0(\theta')).
\label{tbanozeroii}
\end{eqnarray}
where all rapidity integrals here and for the rest of the paper run from
$-\infty$ to $\infty$ unless otherwise labeled. These are the usual TBA
equations for this system \cite{TS,Fowler,pk}, so we conclude that for
$|\alpha|<\pi$ there indeed are no zeroes in $X(\theta)$ or
$Y(\theta)$ for $|Im(\theta)|\le\pi/2$. From (\ref{asyexp}) and
(\ref{tbanozeroii}) it follows that the ground-state energy is
\begin{equation}
E_0=-m\int {d\theta\over 2\pi}\cosh\theta\, \ln(1+X_0(\theta)),
\label{enozero}
\end{equation}
the usual TBA form.

As we will discuss in detail in the next section, $X$ and $Y$ develop
zeroes when $\alpha\ge\pi$. Thus the excited-state energies are given in
terms of $X$ and $Y$ obeying more complicated TBA equations. To find
these, we first note that
$$ \tanh\left({\theta\over 2}-{i\pi\over 4}\right)
\tanh\left({\theta\over 2}+{i\pi\over 4}\right)=1.$$
Note also that in the region $|Im\theta|<\pi/2$, $\tanh((\theta-x)/2)$ is
bounded and its only zero occurs at $\theta=x$. Thus when $X(\theta)$ has
zeroes in this region at $\theta={\rm x}_1,{\rm x}_2,\dots,{\rm x}_J$ (and at $\theta=-{\rm x}_1,
-{\rm x}_2,\dots,-{\rm x}_J$ due to the symmetry under $\theta\to -\theta$)
and $Y(\theta)$ has
zeroes at $\theta={\rm y}_1,{\rm y}_2,\dots,{\rm y}_K$ and their opposites, then
$$
\ln\left[X(\theta) e^{mR\cosh\theta}
\prod_{j=1}^{J} \coth\left({\theta-{\rm x}_j\over 2}\right)
\coth\left({\theta+{\rm x}_j\over 2}\right)\right]
$$
and
$$
\ln\left[Y(\theta)\prod_{k=1}^{K} \coth\left({\theta-{\rm y}_k\over 2}\right)
\coth\left({\theta+{\rm y}_k\over 2}\right)\right]
$$
satisfy the conditions of the Lemma.
Thus the solution of (\ref{tba}) for a given $J$ and $K$ and asymptotic
form (\ref{asy}) obeys the integral equations
\begin{eqnarray}
\ln X(\theta)&=&-mR\cosh\theta +
\sum_{j=1}^J \ln\left[\tanh\left({\theta-{\rm x}_j\over 2}\right)
\tanh\left({\theta+{\rm x}_j\over 2}\right)\right]\nonumber\\
&&\quad +\int {d\theta'\over 2\pi}
{1\over\cosh(\theta-\theta')}\ln\left[(1+e^{i\alpha} Y(\theta'))
(1+e^{-i\alpha} Y(\theta'))\right]
\label{tbamzeroi}\\
\ln  Y(\theta)&=&\sum_{k=1}^K \ln\left[
\tanh\left({\theta-{\rm y}_k\over 2}\right)
\tanh\left({\theta+{\rm y}_k\over 2}\right)\right]
\nonumber\\
&&\quad + \int {d\theta'\over 2\pi}
{1\over\cosh(\theta-\theta')}\ln(1+X(\theta'))
\label{tbamzeroii}
\end{eqnarray}
The energy associated with this solution is then
\begin{equation}
E(\alpha)=-{1\over L}\ln{\cal I}(\alpha)=-m\int {d\theta\over 2\pi}
\cosh\theta\, \ln(1+X(\theta;\alpha))
+ 2\sum_{k=1}^{K} m\cosh({\rm y}_k)
\label{ezero}
\end{equation}

The positions of the zeroes are fixed by
the consistency conditions
\begin{eqnarray}
Y({\rm x}_j\pm i{\pi\over 2};\alpha)&=&-e^{i\alpha} \hbox{ or } -e^{-i\alpha}
\label{firstzero}\\
X({\rm y}_k\pm i{\pi\over 2};\alpha)&=&-1
\end{eqnarray}
which follow immediately from the fact that (\ref{tba}) must still hold
at the locations of the zeroes. We can write these relations
as integral equations
by using (\ref{tbamzeroi},\ref{tbamzeroii}):
\begin{eqnarray}
\pm\alpha -(2M_j+1)\pi&=&
i\sum_{k=1}^K \ln\left[\tanh\left({{\rm x}_j-{\rm y}_k\over 2}+
{i\pi\over 4}\right) \tanh\left({{\rm x}_j+{\rm y}_k\over 2}+
{i\pi\over 4}\right)\right]
\nonumber\\
&&\quad + \int {d\theta\over 2\pi} {1\over\sinh({\rm x}_j-\theta)}
\ln(1+X(\theta))
\label{mzeroi}\\
-(2N_k+1)\pi&=& mR\sinh({\rm y}_k)
\nonumber\\
&&\quad +i\sum_{j=1}^J \ln\left[
\tanh\left({{\rm y}_k-{\rm x}_j\over 2}+{i\pi\over 4}\right)
\tanh\left({{\rm y}_k+{\rm x}_j\over 2}+{i\pi\over 4}\right)\right]
\nonumber\\
&&\quad + \int {d\theta\over 2\pi} {1\over\sinh({\rm y}_k-\theta)}
\ln\left[(1+e^{i\alpha}Y(\theta))(1+e^{-i\alpha}Y(\theta))\right],
\label{mzeroii}
\end{eqnarray}
where the $M_j$ and $N_k$ are integers associated with a given solution.
These integral equations are convenient for numerical solution.
The branch of the logarithm is defined to be between $-i\pi$ and $i\pi$;
redefining this merely corresponds to redefining $M_j$ and $N_k$.
Note that there are no poles in the integrands at $\theta={\rm x}_j$ or
$\theta={\rm y}_k$ because $X({\rm x}_j)=Y({\rm y}_k)=0$.
We will show in the next section that continuity in
$\alpha$ uniquely specifies the $M_j$ and $N_k$ for a given $\alpha$.
Thus by studying how zeroes appear,
one finds $E(\alpha)$ and hence ${\cal I}(\alpha)$ for all $\alpha$.

There is one special case of $\alpha$ where these equations are
solvable in closed form.
This is $\alpha=\pi$, where ${\cal I}$ is Witten's index. As
discussed in \cite{CFIV}, we have $X(\theta)=0$ and $Y(\theta)=1$ for
all $\theta$. This yields the required $E_1=0$ as discussed in section
2.

\section{${\cal I}(\alpha)$ in the massless limit}

In  this section we will analyze the solutions of the TBA equations
in the limit $m\to 0$. We define 
$${\cal X}(\theta;\alpha)=\lim_{m\to 0} X(\theta-\ln(mR/2);\alpha) \qquad
{\cal Y}(\theta;\alpha)=\lim_{m\to 0} Y(\theta-\ln(mR/2);\alpha) 
$$
As discussed in \cite{BLZ}, these functions ${\cal X}$ and ${\cal Y}$ are entire functions of $\theta$ {\bf and} $\alpha$. In this section we
exploit this analyticity to find integral equations which determine
them for all $\alpha$.
This requires showing at which values of $\alpha$ zeroes enter the strip
$|Im\theta|\le\pi/2$ and the corresponding $M_j$ and $N_k$.

The functions  ${\cal X}$ and ${\cal Y}$ obey
\begin{eqnarray}
\ln {\cal X}(\theta)&=&-e^\theta +  \sum_{j=1}^J
\ln\tanh\left({\theta-x_j\over 2}\right)\nonumber\\ &&
\quad +\int {d\theta'\over 2\pi} {1\over\cosh(\theta-\theta')}
\ln\left[(1+e^{i\alpha}{\cal Y}(\theta'))
(1+e^{-i\alpha}{\cal Y}(\theta'))\right] \label{tbazeroi}\\
\ln {\cal Y}(\theta)&=&
\sum_{k=1}^K \ln\tanh\left({\theta-y_k\over 2}\right) +
\int {d\theta'\over 2\pi} {1\over\cosh(\theta-\theta')}\ln(1+{\cal
X}(\theta'))\quad
 \label{tbazeroii}
 \end{eqnarray}
and
\begin{eqnarray}
\pm\alpha -(2M_j+1)\pi&=& i\sum_{k=1}^K \ln\tanh\left({x_j-y_k\over 2}+
{i\pi\over 4}\right)\nonumber\\
&&\quad +\int {d\theta\over 2\pi} {1\over\sinh(x_j-\theta)}
\ln(1+{\cal X}(\theta))\label{zeroi}\\
-(2N_k+1)\pi&=& e^{y_k} +i\sum_{j=1}^J \ln\tanh\left({y_k-x_j\over 2}+
{i\pi\over 4}\right)\nonumber\\
&&\quad+\int {d\theta\over 2\pi} {1\over\sinh(y_k-\theta)}
\ln\left[(1+e^{i\alpha}{\cal Y}(\theta))(1+e^{-i\alpha}{\cal Y}(\theta))\right].
\label{zeroii}
\end{eqnarray}
Roughly speaking, these are ``half'' the solutions of the massive case.
As $mR\to 0$, the functions $X(\theta)$
and $Y(\theta)$ are constant in the region $|\theta|\ll
\ln(2/mR)$; this constant region becomes the asymptotic region
$\theta\to-\infty$ in ${\cal X}(\theta)$ or ${\cal Y}(\theta)$.

Continuity requires that as we vary $\alpha$, there are only two
ways for a new zero to enter the strip $|Im\theta|<\pi/2$. The first
is for a single zero to enter at $\theta=-\infty$. The second is for a
pair of zeroes to enter, one at $\theta=z+i\pi/2$, and the other at
$\theta=z-i\pi/2$. We adopt the conjecture of \cite{BLZ} that all
zeroes in the strip occur at real values of $\exp(2\theta/3)$.
and therefore are on the real-$\theta$ axis.
New ones must enter the strip at
$\theta=-\infty$. (The only exception to this statement is at
$\alpha=\pi$, where ${\cal X}(\theta)=0$ for all $\theta$.)

The first step to determining which zeroes are present at a given value
of $\alpha$ is to solve the equations (\ref{tbazeroi},\ref{tbazeroii})
in the limits $\theta\to \pm\infty$.
This enables us to find the possible values of
$\alpha$ where a zero enters or leaves at $\theta=-\infty$. In
subsequent steps we will show which of these places they do
enter and leave (it turns out any time one can enter at $- \infty$, it
does). Since ${\cal X}(\theta)$ and ${\cal Y}(\theta)$ approach
constants as $\theta\to-\infty$, we can do the integrals in
(\ref{tbazeroi}) and (\ref{tbazeroii}), yielding
\begin{eqnarray}
\relax
[{\cal X}(-\infty)]^2 &=& 1 + 2\cos\alpha \,{\cal Y}(-\infty) +
 [{\cal Y}(-\infty)]^2
\label{xinf}\\
\relax
[{\cal Y}(-\infty)]^2 &=& 1 +  {\cal X}(-\infty)
\label{yinf}
\end{eqnarray}
Requiring that the solution be positive at $\alpha=0$ fixes
\begin{equation}
{\cal X}(-\infty)= {\sin(\alpha)\over \sin(\alpha/3)} \quad\quad
{\cal Y}(-\infty)= 2\cos(\alpha/3).
\label{minusinf}
\end{equation}
In general, this means at $\alpha=(3r+1)\pi$ and
$\alpha=(3r+2)\pi$, the number of zeroes (including those at $\infty$) in
${\cal X}$ must change by $1(mod 2)$.
Similarly, at $\alpha=(3r+3/2)\pi$, the number of zeroes in
${\cal Y}$ must change by $1(mod 2)$.
For example,
as $\alpha\to 9\pi/2$, either an existing zero in ${\cal
Y}(\theta)$ returns to $\theta\to-\infty$ and goes away, or a new one
enters. (We rule out pathologies like triple zeroes in the appendix.)

Other useful pieces of information are the values as $\theta\to \infty$,
which are
\begin{equation}
{\cal X}(\theta\to\infty)\to 2\cos(\alpha/2) e^{-\exp(\theta)}
\qquad {\cal Y}(\infty)= 1\ .
\label{plusinf}
\end{equation}
The equation (\ref{tbazeroi}) only requires that that the
coefficient in (\ref{plusinf}) be $\pm \sqrt{4\cos^2(\alpha/2)}$, but
analyticity in $\alpha$ fixes the above result. For example, we see
that a zero must appear in ${\cal Y}(\theta)$ at $\alpha=3\pi/2$, since
${\cal Y}(-\infty)<0$ and ${\cal Y}(\theta\to\infty) \to 1$ for
$3\pi/2 <\alpha <9\pi/2$.
Similarly, there must be a zero in $\cx$ when
$2\pi <\alpha <3\pi$.
Knowing the sign of the exponentially small term in (\ref{plusinf})
lets us see when zeroes go to $+\infty$. Because the
asymptotic behavior of ${\cal X}(\theta)$ at
large $\theta$ changes from
$0^\pm$ to $0^\mp$ at $\alpha=\pi,3\pi,5\pi,...$, continuity requires that
one of the zeroes $x_j$ goes to $\infty$ and
stays there. Once it is at $\infty$, it no longer needs to obey
(\ref{zeroi}), so it cannot move back to finite values.

To find out what happens at higher values of
$\alpha$ we must study the equations for the zeroes (\ref{zeroi}) and
(\ref{zeroii}) in more detail.  These equations simplify at the values
of $\alpha$ where a zero is entering. For example, a zero $y_k$
enters at $-\infty$, and in this limit
the log terms become constants. Moreover, 
${\cal Y}(-\infty)=0$ when a zero enters, so the integrand
in (\ref{zeroii}) vanishes as
$x_j\to-\infty$ at these values of $\alpha$.
Thus when a zero $y_k$ enters or leaves at $\alpha=(3r+3/2)\pi$
by going to $-\infty$, (\ref{zeroii}) simplifies to
\begin{equation}
(2N_k+1) = J
\label{entery}
\end{equation}
Because of the $e^{y_k}$ term in (\ref{zeroi}),
we see that zeroes in ${\cal Y}$ cannot
go to $+\infty$ like those in ${\cal X}$.

Similarly, for a zero $x_j$ entering (or potentially
leaving) ${\cal X}$ at $\alpha=(3r+1)\pi$ or
$(3r+2)\pi$ by going to $-\infty$, (\ref{zeroi}) simplifies to
\begin{equation}
\alpha -(2M_j+1)\pi =-\pi K
\label{enterx}
\end{equation}
where the $+$ sign in (\ref{zeroi}) must be chosen
because as $x_j$ increases
from $-\infty$, the terms on the right-hand side increase.
(This can be checked by plugging in
the expansion ${\cal Y}(\theta)\sim 2\cos(\alpha/3) + A\exp(2\theta/3)
+ B\exp(4\theta/3)+\dots$ valid for $\theta\to-\infty$ into
the relation (\ref{firstzero}); $A$ is given in
(\ref{pert}) below.) We show in the appendix that
choosing the
$+$ sign gives the correct contribution to the energy.
For a zero $x_j$ going to $+\infty$ at  $\alpha=(2r+1)\pi$, one has
\begin{equation}
\pm\alpha -(2M_j+1)\pi =0\ .
\label{exitx}
\end{equation}
Thus we see that when a zero $x_j=-\infty$ enters at $\alpha=\alpha_j$,
the value $x_j$  will
go to $\infty$ at $\alpha_j+K_j\pi$, where $K_j$ is the number of zeroes
in ${\cal Y}$ at $\alpha_j$. This phenomenon is displayed in fig.\ 1,
where we plot the locations of various zeroes of ${\cal X}$ as
functions of $\alpha$.
\begin{figure}
\begin{center}
{\includegraphics[scale=0.6]{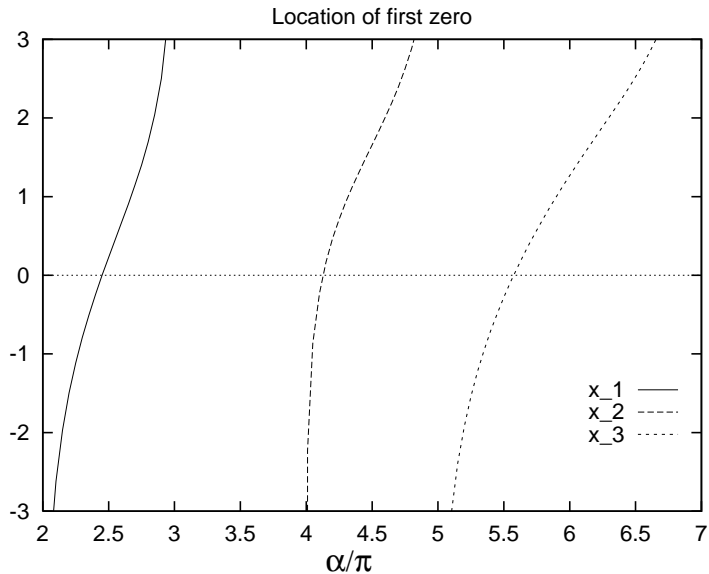}}
\caption{The location of the first zero in ${\cal X}$ as a function
of $\alpha$}
\end{center}
\end{figure}

Now we can map out how at which values of $\alpha$ the zeroes enter 
at $-\infty$, and at which values the zeroes in ${\cal X}$ approach
$+\infty$. For $\alpha<\pi$, there are no zeroes, and the original
TBA equations (\ref{tbanozeroi},\ref{tbanozeroii}) apply. The
situation at $\alpha=\pi$ is atypical because of the special property
that ${\cal X}(\theta)=0$ for all $\theta$. Because $K=0$ here,
(\ref{enterx}) and (\ref{exitx}) are the same, the zero with
$M_1=0$ enters at $-\infty$ and goes straight to $+\infty$. 
The effect of a single zero $x_1$ at infinity is to ensure that ${\cal
X}(\theta)<0$ for all $\theta$.
Thus the TBA equations for
$\pi<\alpha\le 3\pi/2$ are almost the same as the  ordinary TBA
equations (\ref{tbanozeroi}) and (\ref{tbanozeroii}): the only
difference is that $\ln[{\cal X}]$ is replaced by $\ln[-{\cal X}]$.

A zero $y_1$ enters ${\cal Y}$ at $\alpha=3\pi/2$. Because of the zero
$x_1$ already at infinity, $J=1$ here, and we have $N_1=0$. At
$\alpha=2\pi$, ${\cal X}$ changes sign at $-\infty$ but not at
$+\infty$. Therefore, a zero $x_2$ enters at $-\infty$, and since $K=1$
here, $M_2=1$. This zero then goes to $\infty$ at $\alpha=3\pi$,
changing the $\ln[-{\cal X}]$ back to $\ln[{\cal X}]$;
effectively there is no zero in ${\cal X}$ for $3\pi\le\alpha\le 4\pi$.
At $\alpha=4\pi$, a new one must appear because ${\cal X}(\theta)$
changes sign at $\theta\to-\infty$ but not near $\infty$. Still $K=1$,
so this new zero $x_3$ must have $M_3=2$. As we said above, at
$\alpha=9\pi/2$, either a new zero $y_2$ appears, or the old one $y_1$
goes away. The former must happen because $J$ has changed from when
$y_1$ entered; a zero with $N_1=0$ can no longer satisfy (\ref{entery})
and hence can not go back to $-\infty$. The new zero $y_2$ has
$N_2=1$. The zero $x_3$ with $M_3=2$ disappears at $\alpha=5\pi$, but
again a new one must appear. This new zero has $M_4=3$, and since
$K=2$ here, it will go to infinity at $\alpha=7\pi$.

This pattern repeats $mod 6\pi$.
New zeroes enter ${\cal X}$ at $\alpha=(6r+1)\pi,(6r+2)\pi,(6r+4)\pi$ and
$(6r+5)\pi$ with $M_j=4r,4r+1,4r+2$ and $4r+3$, respectively.
Zeroes enter ${\cal Y}$ at $\alpha=(6r+3/2)\pi$ and $\alpha=(6r+9/2)\pi$
with $N_k=2r$ and $2r+1$, respectively. Zeroes $x_j$ go to $\infty$
at $\alpha=(6r+1)\pi, (6r+3)\pi$ and $(6r+5)\pi$. 
Thus when $\alpha$ goes up by $6\pi$, $J$ increases by $4$, $K$ increases
by $2$, and the number of zeroes at infinity goes up by $3$. To summarize,
\begin{eqnarray}
x_{2r-1}& \hbox{ enters at } &\alpha=(3r-2)\pi\nonumber\\
x_{2r}& \hbox{ enters at } &\alpha=(3r-1)\pi\nonumber\\
x_j& \hbox{ goes to }\infty \hbox{ at } &\alpha=(2j-1)\pi\nonumber\\
y_{k}& \hbox{ enters at } &\alpha=(3k-3/2)\pi\nonumber
\end{eqnarray}
while
\begin{eqnarray}
M_j&=&j-1\nonumber\\
N_k&=&k-1\nonumber
\end{eqnarray}
All of this analysis is completely consistent with numerical solutions
of the equations (\ref{tbazeroi},\ref{tbazeroii},\ref{zeroi},\ref{zeroii}).
We present some plots of the functions ${\cal X}(\theta)$ and ${\cal Y}
(\theta)$ for various values of $\alpha$ in figs. 2 and 3.

\begin{figure}
\begin{center}
{\includegraphics[scale=0.6]{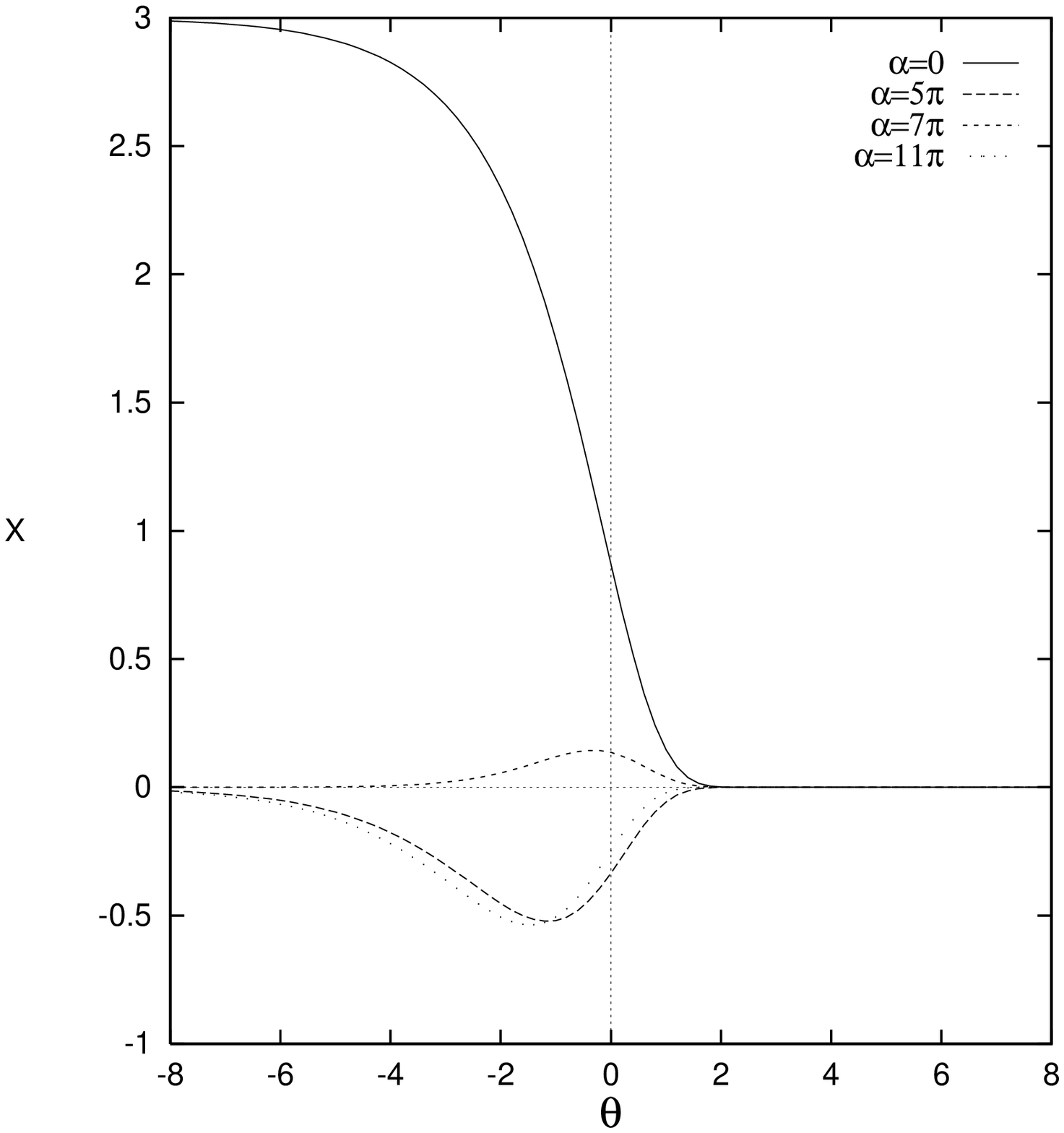}}
\caption{${\cal X}(\theta)$ for several values of $\alpha$}
\end{center}
\end{figure}

\begin{figure}
\begin{center}
{\includegraphics[scale=0.6]{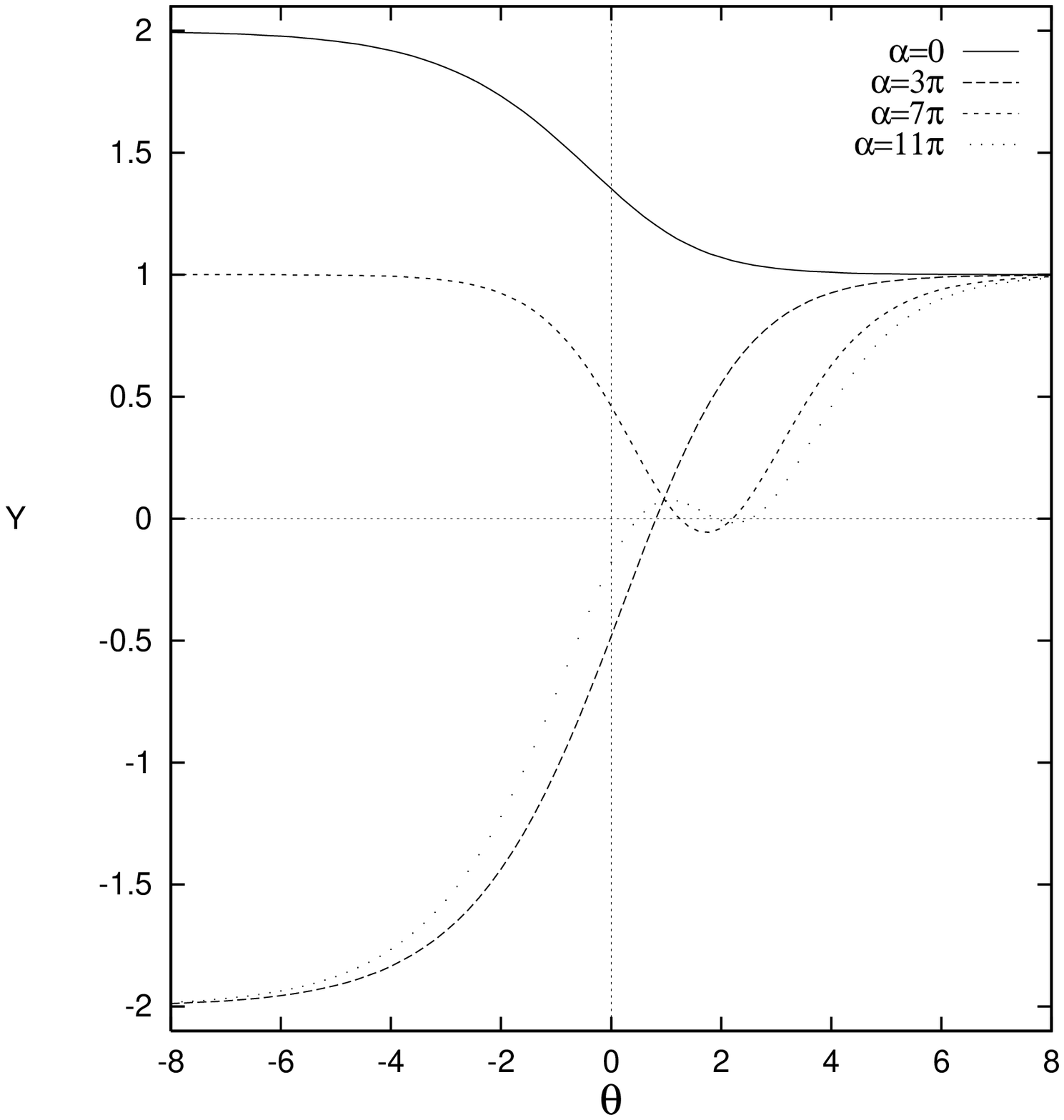}}
\caption{${\cal Y}(\theta)$ for several values of $\alpha$}
\end{center}
\end{figure}

Another very strong check of the results comes from comparison
to perturbative results for $\theta$ near $-\infty$.
In \cite{FLS,BLZii}, it is shown that
\begin{equation}
{\cal Y}(\theta)=2\cos(\alpha/3) 
-{1\over 2^{4/3}}{[\Gamma(-{1\over 3})]^2\Gamma({2\over 3}) \over
\Gamma({1\over 3}+{\alpha\over 3\pi})
\Gamma({1\over 3}-{\alpha\over 3\pi})}
 e^{2\theta/3} +\dots
\label{pert}
\end{equation}
(this corresponds to the coupling $g=2/3$ in \cite{FLS}).
This calculation is done by conformal perturbation theory and
is completely independent of the TBA. In fact, it is not known
how to extract this analytically from the TBA equations. However,
we have checked that (\ref{pert}) is indeed satisfied by numerically
solving the TBA equations for numerous values of $\alpha$.

This scenario is not the unique way of adding zeroes to the functions
consistent with the relations (\ref{enterx},\ref{exitx},\ref{entery}),
but it is the simplest. All the numerical evidence favors this scenario,
as we have seen in the plots.
In the appendix we calculate the energy, and show that for it to agree
with the required value (\ref{Eal}), many other possibilities
can be ruled out. However, there seems to be a strange possibility that
a zero could enter at $+\infty$ with a minus sign in (\ref{tbazeroi}). 
The analysis of the appendix does not rule out the possibility that at
$\alpha=(6r+1)\pi$ or $(6r+5)\pi$, additional zeroes simultaneously
enter at $+\infty$ and $-\infty$. While we cannot rigorously rule this
out, we adopt the assumption of \cite{BLZ} that zeroes enter only
from $-\infty$, and our
numerical work supports this assumption.

These results are also completely in agreement with another
conjecture of \cite{BLZ} concerning where zeroes enter their quantum
monodromy operator. In our case, this agrees perfectly with
our result that zeroes
should enter ${\cal Y}$ at $\alpha=(3r+3/2)\pi$.

\section{Excited-state energies for arbitrary mass}

Once a mass is included, the functions $X(\theta;\alpha)$ and
$Y(\theta;\alpha)$ remain analytic in $\theta$ for a given value of
$\alpha$.  They also are continuous in $m$ for fixed
$\alpha$. However, thay are no longer analytic in $\alpha$, because
they have discontinuities at certain values. We will utilize the
results of the previous section to find which zeroes are present in
the TBA equations with a mass (\ref{tbamzeroi},\ref{tbamzeroii}). In
an appendix, we perform a similar computation for a free Dirac
fermion.

The functions $X(\theta)$ and $Y(\theta)$ are even in $\theta$. Thus
zeroes appear in pairs $({\rm x}_j,-{\rm x}_j)$ and $({\rm y}_k,-{\rm y}_k)$. 
If ${\rm x}_j$ satisfies (\ref{mzeroi}) 
with $M_j$, then $-{\rm x}_j$ solves it with $-M_j$.
Because $\lim_{m\to
0}X(\theta-\ln (2/m);\alpha)={\cal X}(\theta;\alpha)$  and likewise for
$Y$, the equations
$(\ref{mzeroi},\ref{mzeroii})$ smoothly turn into
$(\ref{zeroi},\ref{zeroii})$ as one takes the limit.
Since the $M_j$
and $N_k$ are integers and the other pieces of
$(\ref{mzeroi},\ref{mzeroii})$ are continuous in ${\rm x}_j$ and ${\rm y}_k$,
the effect of turning on a very small mass is to shift
the location of the zeroes and not change the $M_j$ or $N_k$. So for
$m$ small enough, the zero structure is identical to that discussed in
the previous section. Instead of a zero entering at $-\infty$, we
have a pair which enters at $\pm i\pi/2$. As before, we
discount the possibility that a quartet of zeroes enters at conjugate
values ($x+i\pi/2$,$x-i\pi/2$,$-x+i\pi/2$,$-x-i\pi/2$) as $mR$ is
increased. We see no numerical evidence for a quartet entering.

Our strategy in this section will be to fix $\alpha$ and start at $mR$
small where we know the zeroes and the $M_j$ and $N_k$. Once $m$ is
large enough, there is no guarantee that a given ${\rm x}_j$ or ${\rm
y}_k$ will remain real or even stay in the strip
$|Im(\theta)|\le\pi/2$.  For small $mR$, a zero of $X(\theta)$ at
$\theta={\rm x}_j$ corresponds to a zero of ${\cal X}(\theta)$ by the
shift ${\rm x}_j\approx x_j - \ln(mR/2)$.  Thus the initial tendency
as $mR$ is increased is for the pair of zeroes to approach the
origin. They can meet at the origin and move off along the imaginary
axis. Eventually, the pair can reach $\theta +i\pi/2$ and $\theta
-i\pi/2$ and thus not enter into the TBA equations any more.  This is
the only way they can go away: zeroes must appear in conjugate pairs,
and we cannot have ${\rm x}_r={\rm x}_s$ because $M_r\ne M_s$ when
$r\ne s$ (and likewise for ${\rm y}_k$). Thus two zeroes cannot meet
and go off into the complex plane except at the origin.

If two zeroes meet
at the origin, then equations (\ref{mzeroi},\ref{mzeroii}) admit a solution ${\rm x}_j=0$ or ${\rm y}_j=0$.
Because the integral vanishes when ${\rm x}_j=0$,
and $\tanh(z+i\pi/4) \tanh(-z +i\pi/4)=-1$,
for ${\rm x}_j=0$ (\ref{mzeroi}) reduces to
$$\alpha-(2M_k+1)\pi=-\pi K.$$
This is identical
to the condition (\ref{enterx}) for a zero ${\rm x}_j$ to enter at
$-\infty$.  At values of $\alpha$ where (\ref{enterx}) is true,
the zero isn't there yet (it's still at $-\infty$). Thus there is no
value of $\alpha$ where ${\rm x}_j=0$, and it is impossible for a zero in
${\rm x}_j$ to leave at any value of $mR$. Of course, (${\rm x}_j,-{\rm x}_j$)
still moves to ($\infty,-\infty$)
at the same values of $\alpha$ as in the massless case,
because the equations for massive and massless are equivalent
when $\theta$ is large.

It is possible for a zero ${\rm y}_k$ to leave at some value of $mR$.
If ${\rm y}_k$ to be zero, then
$$(2N_k+1)=J.$$
This is the same as (\ref{entery}). However, as opposed to (\ref{enterx}),
this can be satisfied at values of $\alpha$ other than the value $\alpha_k$
where the zero ${\rm y}_k$ entered. As long as $J$ hasn't changed from its value
at $\alpha_k$, a pair of zeroes can meet at the origin. Zeroes ${\rm y}_k$ enter
at the values $\alpha_k=(3k-3/2)\pi$. The next time the
value of $J$ changes is at $\alpha=(3k-1)\pi$. Thus it is possible for
zero ${\rm y}_k$ to leave when $(3k-3/2)\pi <\alpha\le (3k-1)\pi$.
Our numerical evidence suggests that in this range it does in fact leave at
some value of $mR$. To illustrate this, in fig.\ 4
we present a plot of $Y(\theta)$ for $\alpha=2\pi$
for various values of $mR$. At $mR \approx .2$,
the zeroes meet at the origin and thus $Y(\theta)>0$
for all real $\theta$. Then, for $mR\approx .35$, they
leave the strip $|Im\theta|\le \pi/2$ and disappear from the TBA equations
altogether. We see that $Y(\theta)$ is continuous in $mR$ for this
fixed value of $\alpha$.

\begin{figure}
\begin{center}
{\includegraphics[scale=0.7]{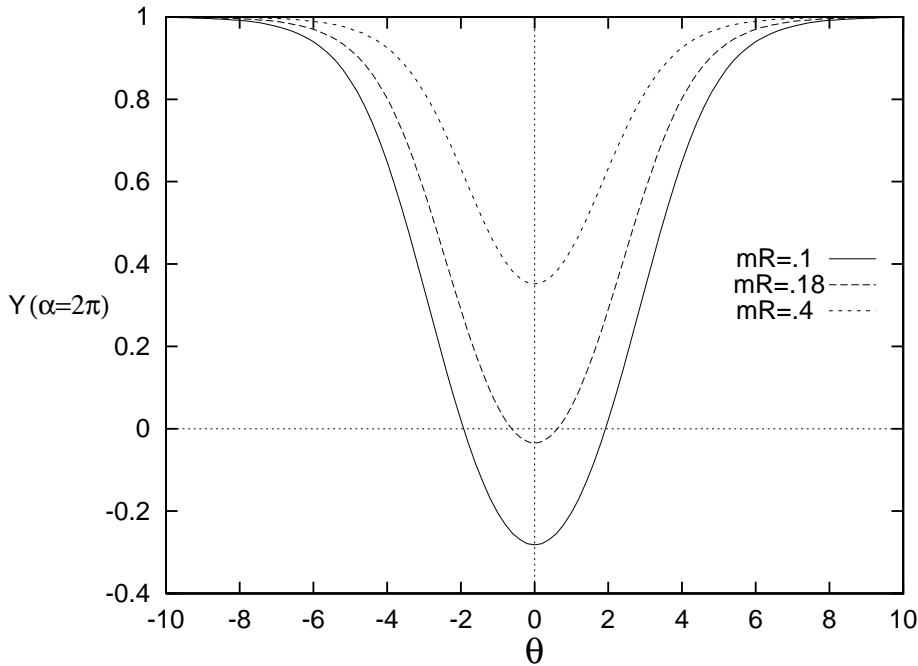}}
\caption{$Y(\theta)$ at $\alpha=2\pi$ for several values of $mR$}
\end{center}
\end{figure}

This observation lets us see at which values of $\alpha$ the functions
are discontinuous when $mR>0$. At $\alpha=2\pi$, when $mR>.4$, there is
no zero in $Y(\theta)$. However, for $\alpha>2\pi$, there is a zero for
any value of $mR$. This zero enters discontinuously, as we can see by
looking at fig.\ 5, where ${\cal Y}(\theta)$ is
plotted at $mR=1$ for $\alpha=2\pi$ and  $2.0000001\pi$.
The functions are discontinuous even at $mR<.2$. There are similar
discontinuities at the values $\alpha=(3K-1)\pi$.

\begin{figure}
\begin{center}
{\includegraphics[scale=0.7]{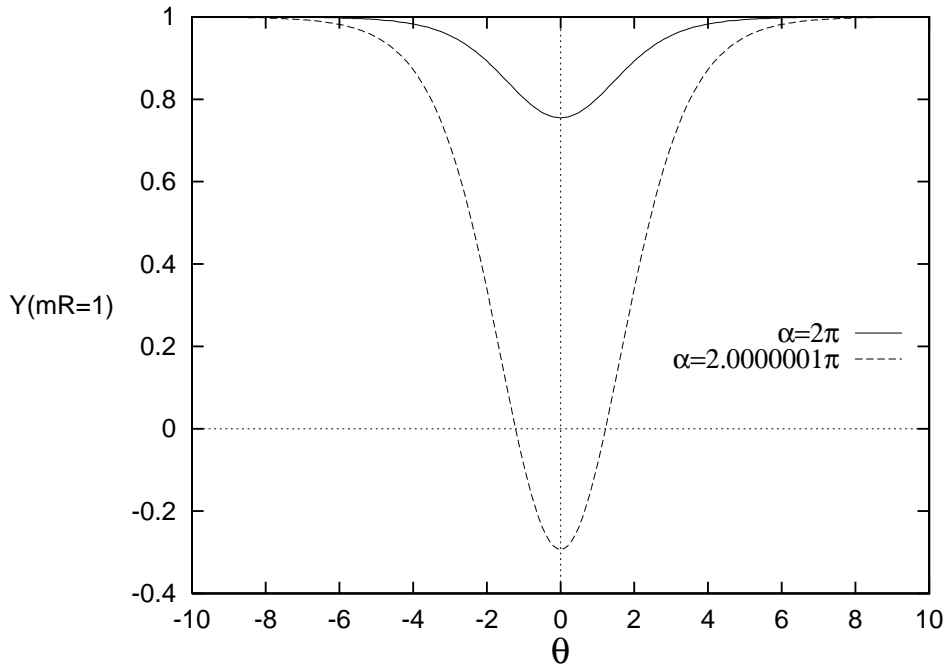}}
\caption{$Y(\theta)$ at $mR=1$ for $\alpha=2\pi$
and $2.0000001\pi$}
\end{center}
\end{figure}

Since the entire analysis of section 4 relied on continuity in
$\alpha$, let us reiterate how we found these discontinuities. In
section 4, we utilized the fact that the functions are continuous in
$\alpha$ in a special limit $mR\to 0$. This enabled us to find the
which zeroes were present as a function of $\alpha$. Then we fix a
value of $\alpha$. For a given $\alpha$, the zeroes move around
continuously as a function of $mR$, so the zero structure is the same
for small enough values of $mR$, and we can thus exploit the $mR\to 0$
results. None of this precludes the possibility that if we fix $mR$
and vary $\alpha$, the functions can change discontinuously. In
appendix B, we show that these discontinuities arise even for a free
Dirac fermion.

By examining the $mR\to\infty$ limit, we see that at
$\alpha=(3K-1)\pi$, the energy shifts from that of a $2(K-1)$-particle
state to a $2K$-particle state. The TBA equations are easily solvable
in this limit, where $X\to 0$ and $Y\to 1$.
The only contribution to the energy (\ref{ezero}) comes
directly from the zeroes ${\rm y}_k$. The integral in
(\ref{zeroii}) goes to zero in the large-$mR$ limit,
and the $mR\sinh({\rm y}_k)$ term must remain finite for there to be a
solution. Thus ${\rm y}_k\to 0$ as
\begin{equation}
\lim_{mR\to\infty} {\rm y}_k \to {(J-2k+1)\pi\over mR},
\label{limy}
\end{equation}
where we used the fact that $N_k=k-1$. In this limit the locations
of the zeroes have the nice interpretation as the rapidities of
the particles in the state.
Because there are $K$ zeroes for large $mR$ when
$(3K-1)\pi<\alpha\le (3K+2)\pi$, we have
\begin{equation}
\lim_{mR\to\infty} E \to 2Km.
\end{equation}
This corresponds to inserting a $2K$-particle state at the each end of
the cylinder. As the cylinder radius $R$ gets large, the particles are
far apart so the energy approaches the constant value $2Km$.
The first
correction is proportional to $1/(mR^2)$, as it should be for a collection
of massive particles far apart: the momentum $p\propto 1/R$, and
for each particle $E=\sqrt{p^2+m^2}\sim m + {\cal O}(1/(mR^2))$.
All other terms in this expansion depend on the interactions between the
particles; the full $E$ gives the interaction energy non-perturbatively.
One can compute the corrections perturbatively by using methods described
in \cite{KM}. In fig.\ 6, we plot $E_3$ as a function of $mR$.
\begin{figure}
\begin{center}
{\includegraphics[scale=0.6]{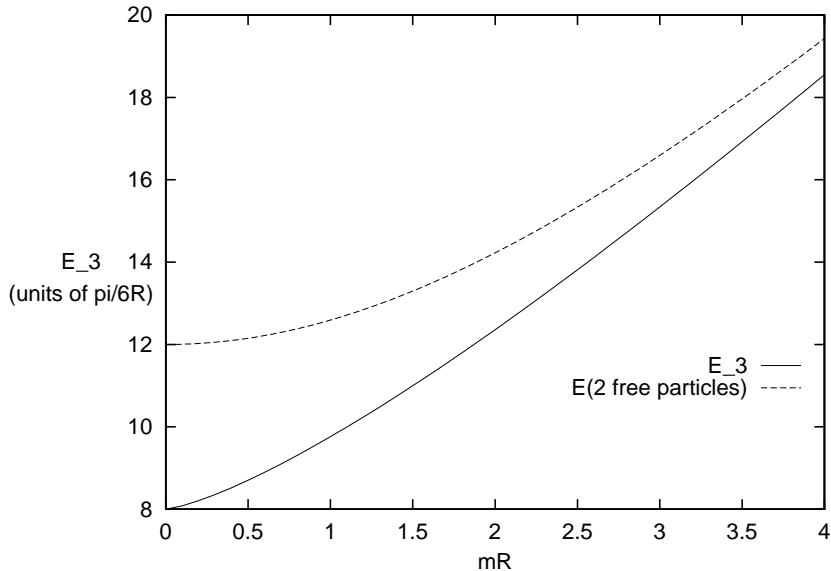}}
\caption{\protect$E_3$ at $\alpha=3\pi$ as a function of $mR$. As $mR$ gets large, $E_3\approx 2\protect\sqrt{m^2+(\pi/R)^2}$, the
energy of two free particles of momenta $\pm\pi/R$.}
\end{center}
\end{figure}

Let us examine $E_3$ and $E_5$ more closely. Both are two-particle
states. There are two kinds of particle, the soliton and antisoliton.
When rewritten in the appropriate $N$=2 language, soliton number is
proportional to fermion number, and because they are a doublet under
the supersymmetry, they must saturate the Bogomolny bound \cite{pk}.
Since the states $|n\rangle$ have zero fermion number, the states
$|3\rangle$ and $|5\rangle$ each consist of a soliton and an
antisolition. From (\ref{limy}) we see
that the individual momenta of the particles in the large $R$ limit are $\pm
\pi/R$ in $|3\rangle$, and $\pm 2\pi/R$ in $|5\rangle$. As discussed in
\cite{dorey}, one can presumably obtain the energy of two-particle
states with higher momenta by using different values of $M_j$ and
$N_k$. We have
checked numerically that there are such solutions of the TBA
equations, but have not pursued this
line of investigation any further.

\section{Conclusion}

One application of this work is to the system of two-dimensional circular
polymers. As discussed in \cite{FS}, partition functions calculated in
this model are related to scaling functions for the number of closed
self-avoiding random walks on a cylinder. The world-lines of the
particles of this model correspond to the random walk, or equivalently
the polymer. A single closed polymer will correspond to a two-particle
state. Thus when we insert a two-particle state on each end of the
cylinder, it corresponds to a circular polymer stretched from one end to
the other.

In this paper we have calculated various excited-state energies in the
simplest integrable model with $N$=2 supersymmetry. We view this as a
useful exercise, because it is a unitary theory, and because it shows
that generalizing the work of \cite{BLZ,dorey} to models
with non-diagonal $S$ matrices is not prohibitively
complicated. It seems likely that one could calculate the excited-state
energies for any value of the coupling in sine-Gordon in this manner.
Moreover, by choosing different values of the integers $M_j$ and $N_k$,
it seems likely that many more excited-state energies could be obtained.

Our results also yield an interesting set of expectation values in
$N$=2 supersymmetric theories which are independent of $D$-term
perturbations like the ``index'' studied in \cite{CFIV}. That index had
some remarkable properties in that the TBA equations turned out to be
equivalent to a differential equation, Painlev\'e III. Moreover, 
because it depends on the parameters of the theory, it contains
much information about the solitons of the theory. For example, it
found useful application in four-dimensional gauge theories \cite{HM}.
This makes us hopeful that
the results of this paper will also find future application.

\bigskip\bigskip
I would like to thank P. Dorey, K. Intriligator and H. Ooguri for
helpful conversations. This paper was supported by NSF grant no.
9413208.

\appendix
\section{The energy in the massless limit}

In this appendix, we calculate the energy in the $m\to 0$ limit
and verify that it gives the correct answer (\ref{Eal}) with
our values of $M_j$ and $N_k$. This
has the bonus of ruling out some alternative scenarios for the
appearance of zeroes.

We define ${\cal E}(\alpha)\equiv R\lim_{m\to 0}E(\alpha)$. Then
the expression (\ref{ezero}) for the energy yields
\begin{equation}
{\cal E}(\alpha)=-{1\over\pi}\int e^{\theta} L_{\cal X}(\theta)
+2\sum_{k=1} e^{y_k}
\label{calE}
\end{equation}
where $L_{\cal X}(\theta)\equiv \ln(1+X(\theta))$. Similarly, we define
$L_{\cal Y}(\theta)\equiv \ln[(1+e^{i\alpha}{\cal Y}(\theta))
(1+e^{-i\alpha}{\cal Y}(\theta))]$. To simplify this, we use a
well-known TBA trick. We take the derivative with respect to $\theta$
of (\ref{tbazeroi}) and use this to eliminate the $e^{\theta}$ in
(\ref{calE}), yielding
\begin{eqnarray}
{\cal E}=2\sum_{k=1}^K e^{y_k} +\int {d\theta\over \pi}
{\delta \cx\over\delta\theta} {1\over \cx}L_{\cal X}(\theta)
-\sum_{j=1}^J\int {d\theta\over \pi}
{1\over\sinh(\theta-x_j)}L_{\cal X}(\theta)\nonumber\\
-\int\int{d\theta\over\pi}{d\theta'\over 2\pi} L_{\cal X}(\theta)
{1\over \cosh(\theta-\theta')} {\delta L_{\cal X}(\theta')\over\delta\theta'}
\nonumber
\end{eqnarray}
The double integral in the final term can be simplified back to a single
integral by using (\ref{tbazeroii}). After integrating by parts, one finds
\begin{eqnarray}
{\cal E}&=&2\sum_{k=1}^K e^{y_k} +\int {d\theta\over \pi}
{\delta \cx\over\delta\theta} {1\over \cx}L_{\cal X}(\theta)
+\int {d\theta\over \pi}
{\delta {\cal Y}(\theta)\over\delta\theta} {1\over {\cal Y}(\theta)}
L_{\cal Y}(\theta)\nonumber\\
&&-\sum_{j=1}^J\int {d\theta\over \pi}
{1\over\sinh(\theta-x_j)}L_{\cal X}(\theta)
-\sum_{k=1}^K\int {d\theta\over \pi}
{1\over\sinh(\theta-y_k)}L_{\cal Y}(\theta)\nonumber\\
&&-{1\over\pi}\ln[({\cal X}(-\infty))^2]\ln[((-1)^K {\cal Y}(-\infty)]
\nonumber
\end{eqnarray}
where we simplified the final term (the surface term from 
the integration by parts) by using (\ref{xinf},\ref{yinf}). The zeroes
at infinity (there are $J_\infty$ of them) do not contribute, so we can
rewrite the sum over $j$ to start at $J_\infty+1$. Notice
we can change variables to rewrite the first two integrals as simple 
integrals, so for example
$$\int {d\theta\over \pi}
{\delta \cx\over\delta\theta} {1\over \cx}L_{\cal X}(\theta)
=\int_{{\cal X}(-\infty)}^{{\cal X}(\infty)} {du \over\pi}{1\over u}\ln(1+u).
$$
We use
the equations for zeroes (\ref{zeroi},\ref{zeroii}) to eliminate the
terms with the ${1/\sinh}$ and the $e^{y_k}$, yielding
\begin{eqnarray}
{\cal E}=&&\int_{{\cal X}(-\infty)}^{{\cal X}(\infty)}
 {du\over \pi} {1\over u}\ln(1+u)
+\int_{{\cal Y}(-\infty)}^{{\cal Y}(\infty)} {dv\over \pi} {1\over v}
\ln\left[(1+e^{i\alpha}v)(1+e^{-i\alpha}v)\right]\nonumber\\
&&-{1\over \pi}\ln[({\cal X}(-\infty))^2]\ln [{\cal Y}(-\infty)(-1)^K]
\nonumber\\
&&
+2JK\pi -2\sum_{k=1}^K(N_k+1)\pi +2\sum_{j=J_\infty+1}^J 
(\pm\alpha-(2M_j+1)\pi)
\label{uvenergy}
\end{eqnarray}
Because the first three terms are always finite and depend only on
${\cal X}(\pm\infty)$ and ${\cal Y}(\pm\infty)$, they are  periodic
under in $\alpha\to\alpha+6\pi$. However, as discussed in section 2, we
know that
\begin{equation}
{\cal E}(\alpha)= {\pi\over 6}\left(1-{\alpha^2\over\pi^2}\right)
\label{eee}
\end{equation}
 and the full answer
is not periodic. This means that zeroes must enter into ${\cal X}$ and
${\cal Y}$ in order so that the last three terms make ${\cal
E}(\alpha)$ continuous in $\alpha$.
This puts strong constraints on
the values of $M_j$ and $N_k$ and at which values of $\alpha$ the zeroes enter.
Despite the fact that it looks like (\ref{uvenergy}) will be discontinuous
when a zero enters,
the conditions (\ref{entery},\ref{enterx},\ref{exitx}) ensure continuity.
This condition of continuity rules out many other conceivable ways zeroes
could enter.

To simplify it further, we take the derivative of ${\cal E}(\alpha)$
with respect to $\alpha$. By using the relations (\ref{xinf}) and
(\ref{yinf}) and their derivatives with respect to $\alpha$,
most of the resulting terms cancel, leaving only
$${\delta {\cal E}\over\delta\alpha}=
-\int_{2\cos(\alpha/3)}^1{dv\over\pi}
{2\sin\alpha\over 1+2\cos\alpha v +v^2} + \sum_{j=J_\infty+1}^J (\pm 1).$$
For $\alpha<\pi$, one can do the integral, yielding $-{\alpha/ 3\pi}$
in agreement with (\ref{eee}).
However at integer values of $\alpha/\pi$, one must carefully check
if there are discontinuities because the denominator and
numerator both can become zero. One sees that
at $\alpha=(6r+2)\pi$ and $(6r+4)\pi$, the integrand is $2\delta(v+1)$
giving discontinuities of $-1$ in $\delta {\cal E}/\delta\alpha$.
Similarly, at $\alpha=(6r+3)\pi$ there are discontinuities of $+1$.
These discontinuities must
be cancelled by the second term. This means a zero with $+\alpha$
in (\ref{zeroi}) must
enter at $\alpha=(6r+2)\pi$ and $(6r+4)\pi$, while at $\alpha=(6r+3)\pi$,
either one with $-\alpha$ should enter, or one with $+\alpha$ should
go to $\infty$, where it does not contribute to (\ref{uvenergy}).
Since we show in section 4 that a single
zero cannot enter at $\alpha=(6r+3)\pi$, the latter possibility
must hold.

These conditions are satisfied by the scenario discussed in section 4,
and we take it as strong evidence that it is correct. The analysis of
this appendix rules out, for example, the possibility that a triple
zero enters at $\alpha=(6r+1)\pi$.

\section{The excited states of a free Dirac fermion}

In this appendix we derive the excited-state energies of a free Dirac
fermion by applying the general methods of this paper. This will
clarify several issues which arose in the interacting case, in
particular the nature of the discontinuity discussed in section 5,
where the energy is discontinuous at particular values of $\alpha$
when the mass is non-zero.

Just like the interacting case, the expectation value of interest is
(\ref{fun}). This corresponds to placing the fermion and antifermion
at opposite (imaginary) chemical potentials. Although (\ref{fun})
looks periodic in $\alpha$, we define the values for $\alpha>\pi$ as
the appropriate continuation of those for $\alpha\le \pi$.  Since the
particles are free, we can evaluate the trace explicitly.
We define a function $T(\theta)$ which satisfies the functional
relation
\begin{equation}
T(\theta +i\pi/2) T(\theta-i\pi/2)= e^{-mR\cosh\theta} + 2\cos\alpha 
+ e^{mR\cosh\theta}.
\label{afunc}
\end{equation}
It follows that $T(\theta+2\pi)=T(\theta)$.
In this non-interacting case, we can find all the zeroes of $T$
explicitly.
The relation (\ref{afunc}) requires that
$$
\exp\left( i\alpha + mR\cosh({\rm t}_n^\pm+i\pi/2)\right) = -1,
$$
where $T({\rm t}_n^{\pm})=0$ and $T(-{\rm t}_n^{\pm})=0$.
Taking the logarithm, we have
\begin{eqnarray}
\alpha+mR\sinh {\rm t}_n^+ &=& -(2n-1)\pi
\label{azeroi}\\
-\alpha+mR\sinh {\rm t}_n^ -&=& -(2n-1)\pi
\label{azero}
\end{eqnarray}
where $n$ is some integer.
Using the Lemma of section 3, one can write an integral
equation for $\ln T$:
\begin{eqnarray}
\ln[T(\theta)]&=&mR\cosh\theta
+\sum_{k} \ln\left[\tanh\left(\frac{\theta-{\rm t}_k}{ 2}\right)
\tanh\left(\frac{\theta+{\rm t}_k}{ 2}\right)\right]\nonumber\\
&&\ +\int \frac{d\theta'}{ 2\pi}
\frac{1}{\cosh(\theta-\theta')} \ln \left[(1+e^{i\alpha}e^{-mR\cosh\theta})
(1+e^{-i\alpha}e^{-mR\cosh\theta})\right]
\label{atba}
\end{eqnarray}
where the sum over $k$ is only over those zeroes which are in the
strip $-\pi/2 \le Im\, {\rm t}_k \le \pi/2 $.

The energy levels are found by the large-$\theta$ asymptotic expansion
as in (\ref{asyexp}) \cite{BLZ}. A bulk term is present here which does not
appear in the supersymmetric case. This term is independent of
$\alpha$, so we won't bother to discuss this subtlety. 
The energy is
\begin{eqnarray}
E(\alpha)&=&E_0(\alpha)
+ 2\sum_{k} m\cosh({\rm t}_k)\\
E_0(\alpha)&=& -m\int \frac{d\theta}{ 2\pi}
\cosh\theta\,  \ln \left[(1+e^{i\alpha}e^{-mR\cosh\theta})
(1+e^{-i\alpha}e^{-mR\cosh\theta})\right]\nonumber
\label{ae}
\end{eqnarray}
where again the sum over $k$ is over only the zeroes in the strip.
Notice that $E_0(\alpha)$ is periodic in $\alpha$, and denotes the
usual Casimir energy of a Dirac fermion.

We now need to decide which zeroes are present for a given value of
$\alpha$. At $\alpha=0$, $E_0$ is indeed the
correct ground state energy for a Dirac fermion, so there are no zeroes
in the strip here. Thus the zeroes ${\rm t}_n^\pm$ must solve
(\ref{azeroi},\ref{azero})
with $n\ge 1$, so that $Im\, {\rm t}_n^{\pm}=\pi$ when
$\alpha=0$. (The relation (\ref{afunc}) only requres that $T({\rm
t}_n^\pm)=0$ \textbf{or}
$T({\rm t}_n^\pm+i\pi)=0$, so it is consistent to require that
$T({\rm t}_n^\pm)=0$
only for $n$ positive. One can find different excited states of
the system by restricting the allowed
$n$ to be a different subset of the integers.)

As discussed in \cite{BLZ}, in the massless limit
${\cal T}(\theta;\alpha)\equiv \lim_{m\to 0} T(\theta-\ln(mR/2);\alpha)$
is analytic in $\theta$ and $\alpha$. The zeroes $t_n^{\pm}$
of ${\cal T}$ must therefore obey
\begin{eqnarray}
\alpha+ e^{t_n^+} = -(2n+1)\pi\nonumber\\
-\alpha+ e^{t_n^-} = -(2n+1)\pi\nonumber
\end{eqnarray}
with $n\ge 1$ for all $\alpha$.
For $\alpha <\pi$, all the $t_n^\pm$ have
imaginary part $i\pi$, and do not contribute to the energy (\ref{ae}).
At $\alpha=\pi$, $t_1^- =-\infty$. As we go past $\alpha=\pi$, this zero
``jumps'' onto the real axis. Even though this seems
discontinuous, one can easily check
that ${\cal T}$ is continuous: the extra term coming
from the zero in (\ref{atba}) is zero when $t_k=-\infty$.
Similarly, at $\alpha=(2K-1)\pi$, the zero $t_K^-$ jumps onto the real
axis and hence enters into expressions for ${\cal T}$ and the energy.
If we were to take $\alpha$ to $-(2K-1)\pi$ instead, then the zero
$t_K^+$ jumps onto the real axis.

A similar phenomena happens in the massive case. At
$\alpha=(2K-1)\pi$, we see from (\ref{azero}) that the zero ${\rm
t}_K^-$ jumps from $i\pi$ to $0$.
This solution reduces to the massless one appropriately in the
limit $m\to 0$. (It is possible to find a continuous solution to
(\ref{azero}) by having
the zeroes $t_K^-$ and $-t_K^-$ just switch places at $\alpha=(2K-1)\pi$,
but this solution is periodic in $\alpha$ and does not reduce to
the correct massless solution.) Thus for $\alpha>(2K-1)\pi$, the
zero ${\rm t}_K^-$ must now must be included in the sum in
(\ref{atba}) and likewise in (\ref{ae}). However, the interesting
distinction between the massless and massive cases is that $T$ and $E$
are \textbf{not} continuous at these values of $\alpha$, as is easily
seen by examination of (\ref{atba}) and (\ref{ae}).

The physical interpretation of this is clear.  For $(2K-1)\pi < \alpha
\le (2K+1)\pi$, the energy is that of a state of $K$ fermions and $K$
antifermions. The zeroes $\pm {\rm t}_k$ with $k\le K$ are the
rapidities of the particles, so that the energy of a given particle is
$m\cosh{\rm t}_k$. The energy still contains the Casimir piece
$E_0(\alpha)$; because the particles are non-interacting it is not
changed by presence of physical particles in the state.
The example treated in the main part of this paper is interacting, so
the energy only decouples into a ``Casimir'' piece and an
``excited-state'' piece in the large-$R$ (dilute) limit. However, we
have seen in this appendix that the discontinuities in the energy and in
the function used to determine it are very natural physically, if
somewhat surprising mathematically.

\clearpage
\renewcommand{\baselinestretch}{1}


\begin{thebibliography}{99}

\bibitem{YY} C.N. Yang and C.P. Yang, J. Math. Phys. 10 (1969) 1115\\
Al.B. Zamolodchikov, Nucl. Phys. B342 (1990) 695
\bibitem{cardy} J.L. Cardy, Nucl. Phys. B275 (1986) 200
\bibitem{me} P. Fendley, Nucl. Phys. B374 (1992) 667, hep-th/9109021
\bibitem{MKM} M.J. Martins, Phys. Rev. Lett. 67 (1991) 419\\
T. Klassen and E. Melzer, Nucl. Phys. B370 (1992) 511
\bibitem{BLZ} V. Bazhanov, S. Lukyanov and A.B. Zamolodchikov,
Nucl. Phys. B489 (1997) 487, hep-th/9607099
\bibitem{dorey} P. Dorey and R. Tateo, 
Nucl.Phys. B482 (1996) 639, hep-th/9607167
\bibitem{KP} A. Kl\"umper and P. Pearce, J. Stat. Phys. 64 (1991) 13;
Physica A183 (1992) 304
\bibitem{doreynew} P. Dorey and R. Tateo, hep-th/9706140
\bibitem{pk} P. Fendley and K. Intriligator, Nucl. Phys. B372 (1992) 533,
hep-th/9111014
\bibitem{CBS} C. Callias, Commun. Math. Phys. 62 (1978) 213\\
R. Bott and R. Seeley, Commun.  Math. Phys. 62 (1978) 235
\bibitem{Witten} E. Witten, Nucl. Phys. B202 (1982) 253
\bibitem{CFIV} S. Cecotti, P. Fendley, K. Intriligator and C. Vafa,
Nucl. Phys. B386 (1992) 405, hep-th/9204102
\bibitem{SS} A. Schwimmer and N. Seiberg, Phys. Lett. 184B (1987) 191.
\bibitem{BCNA} H.W.J. Bl\"ote, J.L. Cardy and M.P. Nightingale, Phys. Rev.
Lett. 56 (1986)742\\
I. Affleck, Phys. Rev. Lett., 56 (1986) 746.
\bibitem{NienDF}B. Nienhuis in {\it Phase Transitions and Critical Phenomena},
ed. by C. Domb and J. Lebowitz, vol. 11, (Academic Press, 1987)\\
V.S. Dotsenko and V.A. Fateev, Nucl.Phys. B240 (1984) 312
\bibitem{LVW} W. Lerche, C. Vafa, N. Warner, Nucl.  Phys. B324 (1989)
427
\bibitem{poly} E.J. Weinberg, Phys. Rev. D24 (1981) 2669\\
A.P. Polychronakos, Nucl. Phys. B283 (1987) 268 
\bibitem{TS} M. Takahashi and M. Suzuki, Prog. Theor.
Phys. 48 (1972) 2187
\bibitem{Fowler} M. Fowler and X. Zotos, Phys. Rev. B25 (1982) 5806
\bibitem{TW} C. Tracy and H. Widom,  Comm. Math. Phys. 179 (1996) 667,
solv-int/9509003
\bibitem{FLS} P. Fendley, F. Lesage and H. Saleur,
J. Stat. Phys. 85 (1996) 211, cond-mat/9510055
\bibitem{BLZii}V. Bazhanov, S. Lukyanov and A.B. Zamolodchikov,
``Integrable Structure of Conformal Field Theory II'',
hep-th/9604044
\bibitem{KM}T. Klassen and E. Melzer, Nucl. Phys. B382 (1992) 441,
hep-th/9202034
\bibitem{FS} P. Fendley and H. Saleur, Nucl. Phys. B388 (1992) 609,
hep-th/9204094
\bibitem{HM} J. Harvey and G. Moore, Nucl. Phys. B463 (1996) 315, hep-th/9510182


\end{thebibliography}
\end{document}